\documentclass[conference]{IEEEtran}
\IEEEoverridecommandlockouts
\usepackage{cite}
\usepackage{amsmath,amssymb,amsfonts}
\usepackage{graphicx}
\usepackage{textcomp}
\usepackage{xcolor}
\usepackage{epstopdf}
\usepackage{subcaption}

\def\BibTeX{{\rm B\kern-.05em{\sc i\kern-.025em b}\kern-.08em
    T\kern-.1667em\lower.7ex\hbox{E}\kern-.125emX}}
\begin{document}

\title{On Low Complexity RLL Code for Visible Light Communication\\
}

\author{Nitin Jain 
and Adrish Banerjee 

\thanks{The authors are with the Department of Electrical Engineering, IIT Kanpur, Kanpur 208016, India. (e-mail: \{nitjain, adrish\}@iitk.ac.in). }
}

\maketitle

\begin{abstract}
Run length limited (RLL) codes are used in visible light communication (VLC) to avoid flicker and to support different dimming ranges such that communication is not affected by the variation in light intensity. In this paper, we propose a low complexity split phase code as RLL code in serial concatenation with the convolutional code as a forward error correcting code (FEC) for VLC. The split phase code can be represented by a two-state trellis-like code which can provide an iterative gain in FEC-RLL serial concatenation. We also use the extrinsic information transfer (EXIT) chart to explain the iterative decoding behavior of the proposed serial concatenated scheme. Furthermore, we use puncturing and compensation symbols to support various dimming range in VLC.
\end{abstract}


\section{Introduction}
In visible light communication (VLC), the free visible light band (400 to 750 THz) is used for supporting communication particularly for short-range along with illumination \cite{b1}. Here the information is transmitted via the intensity control of light emitting diode (LED). VLC has many advantages such as the use of unlicensed spectrum, less complex intensity modulation with direct detection (IM/DD), and potentially large available bandwidth for communication. 

The VLC system should ensure that communication doesn't affect the two main functionalities of light, namely, the flicker mitigation and dimming support \cite{b1}. Run length limited (RLL) line coding technique is used in VLC system before transmitting the data to avoid long runs of 0's and 1's which is the potential cause of flickering and to provide a constant dimming level of 50$\%$. Once the dimming level is fixed to 50$\%$, it can be varied by puncturing some bits and inserting compensation symbols \cite{b1} according to the dimming requirement. In literature, various RLL codes such as Manchester codes, 4B6B code, and 8B10B codes are proposed \cite{b1} for VLC systems. Typically RLL scheme support only hard decoding \cite{b1}, but soft-in-hard-out (SIHO) and soft-in-soft-out (SISO) decoding of RLL codes are also presented in \cite{b5} and \cite{b6} respectively.

Reed-Solomon (RS) codes and convolutional codes (CC) are proposed as FEC codes in \cite{b1} in combination with RLL codes. Other FEC codes proposed in literature include Reed-Muller (RM) codes \cite{b17}, rate-compatible convolutional code \cite{b18}, turbo codes \cite{b14}, low density parity check (LDPC) codes \cite{b15}, and polar codes \cite{b16}. A joint iterative decoding scheme using serial concatenation of FEC-RLL coding is also presented in \cite{b4}. LDPC with constant weight codes (CWC) \cite{b22}, polar codes with puncturing and compensation symbol insertion \cite{b21,b23} is also proposed for supporting various dimming range in VLC.

In this paper, we propose use of ``split phase code'' as inner RLL code in serial concatenation with the outer convolutional code to provide iterative decoding gain. The split phase code can be represented by a two-state trellis like code which corresponds to low decoding complexity. We use the extrinsic information transfer (EXIT) chart \cite{ten2000design,banerjee2001asymmetric}  to analyze the convergence threshold of the proposed system. The system also supports various dimming range and provide better bit error rate (BER) performance. The main contributions of the paper are as follows:
\begin{itemize}
\item We have used split phase code as RLL code for the first time in VLC system. The split phase code can be viewed as a two-state trellis code which corresponds to low decoding complexity than many of the four-state RLL code such as Miller\cite{b4}, BMC\cite{b4} and eMiller\cite{b24} in literature.
\item We have modeled the VLC system as serial concatenated coding structure that uses the convolutional code as outer code and split phase RLL code as inner code. The proposed model allows low complexity iterative decoding algorithm with  $50\%$ dimming.
\item We have used mutual information based EXIT chart to analyze the iterative decoding performance of proposed VLC system. We have shown that we get better BER performance than many other state of the art coding scheme for VLC.
\item The proposed scheme allows any required level of dimming support.
\end{itemize}

The paper is organized as follows. First, we discuss some past literature on RLL codes in section II, followed by the description of a split phase code and system model in section III. In Section IV, we describe the iterative decoding algorithm and visualization of the decoding trajectory using the EXIT chart. Bit error rate (BER) performance results are presented in section V, followed by the concluding remarks in section VI.

\section{RLL codes: Past work}

Manchester code is a rate, $R=\frac{1}{2}$ RLL code that offers perfect 50$\%$ dimming at the output irrespective of the input bit because each input bit is mapped to one high-level bit `1' and one low-level bit `0'. But this coding scheme is memory independent due to which, there is no trellis-like structure for this code, and hence this cannot be used as in iterative decoding for FEC-RLL serial concatenation coding scheme. 

RLL codes such as bi-phase mark code (BMC) and Miller \cite{b4} (code rate, $R=\frac{1}{2}$) can be used in serial concatenation with a convolutional code and can be decoded iteratively. However, both BMC and Miller do not provide perfect 50$\%$ dimming at the output for individual input bit. Authors in \cite{b4} have proposed the concept of super-symbol to evaluate the dimming percentage, rather than at the individual bit level. They have shown that if the super-symbol size is arbitrarily large, one can achieve perfect dimming of 50$\%$. 

The look-up table (LUT) based coding such as 4B6B (or 8B10B) is also used as RLL coding \cite{b1}, and it provides perfect 50$\%$ dimming irrespective of the input sequence. The code rate is $R=\frac{2}{3}$ for 4B6B scheme. The authors in \cite{b3} proposed SISO decoding of 4B6B codes by exchanging extrinsic information with the FEC module using iterative decoding. However, this scheme doesn't provide much improvement with iterations. 

Recursive unity rate code (URC) having the generator polynomial $G(D)=\frac{1}{1+D}$ was proposed in \cite{b3}, to replace the classic RLL codes and provide fluctuating dimming level centered around 50$\%$. \cite{b3} shows that as M ($\#$ bits) increases, the deviation around the dimming value of 50$\%$ decreases. However, there are some typical sequences which will give dimming value far away from 50$\%$, independent of M. 


New Miller code named as eMiller is also proposed as RLL code in \cite{b24} for VLC. This coding scheme, when used in concatenation with convolutional code performs better in the first iteration but doesn't provide any gain with subsequent iterations just like the Manchester code.

\section{Split phase code $\&$ System Model}

We propose the use of the split phase code as a new RLL code in this section. Split phase code has many advantages such as low decoding complexity and perfect dimming of 50$\%$ for each input bit without considering the concept of super-symbol. It also avoids flicker because of the run length of two.

\begin{figure}[t]
\centerline{\includegraphics[scale=0.08]{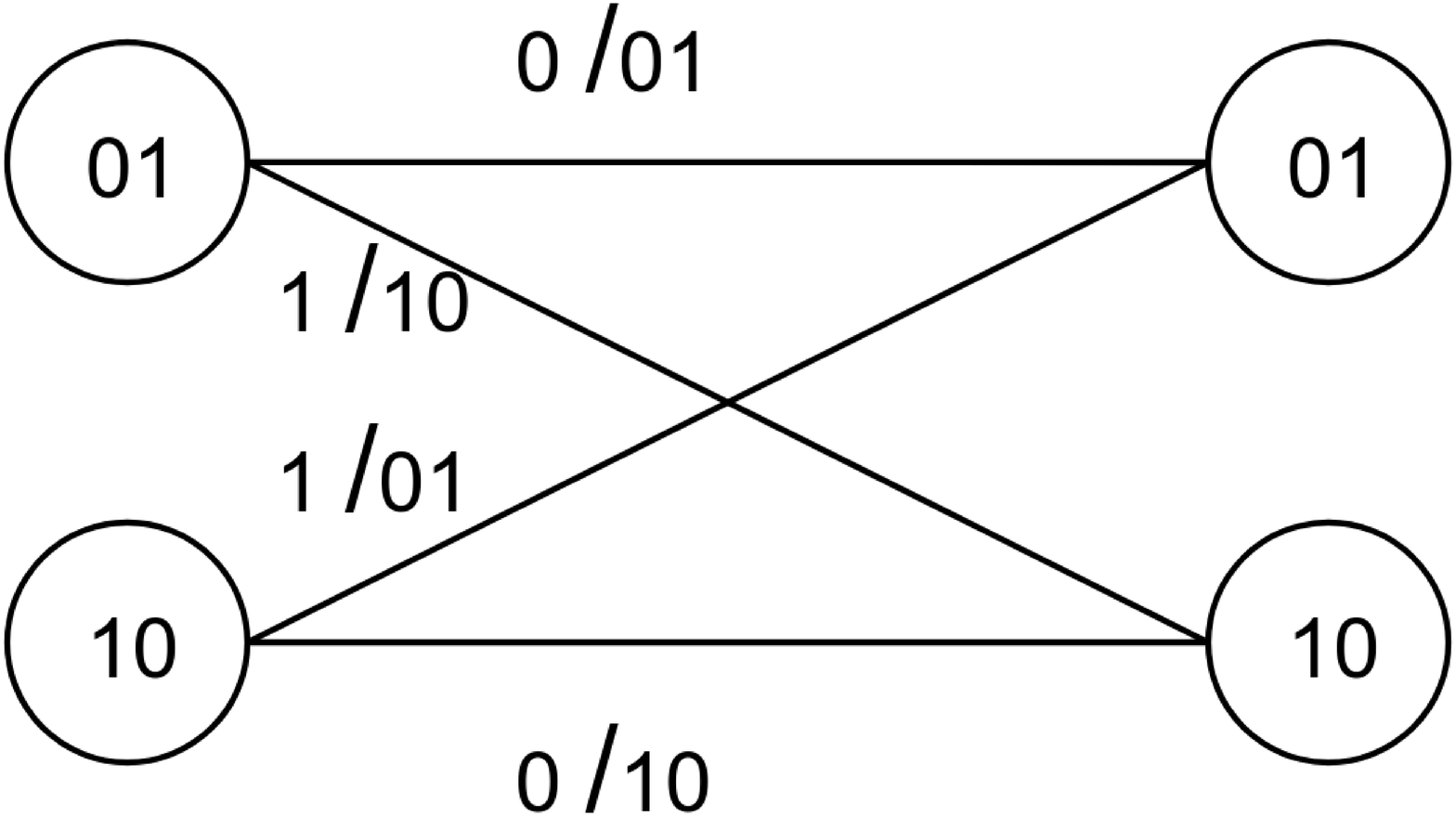}}
\caption{State Diagram of Split Phase code}
\label{fig:splitphase}
\end{figure}

The trellis diagram of split phase code is shown in Fig. \ref{fig:splitphase}. The code rate is $R=\frac{1}{2}$, as it gives two output bits corresponding to each input bit (output bits are same as state bits). This scheme provides a perfect 50$\%$ dimming because each input bit is mapped to one high-level bit `1' and one low-level bit `0'. 
The split phase code is memory dependent, and one can use iterative decoding for split phase code, unlike the Manchester code. The split phase code is less complex than the BMC/Miller code as its underlying structure is a two-state trellis diagram. Table \ref{table:tab2} summarizes the comparison of RLL codes proposed in the literature with the split phase RLL code.  

The system model of the VLC system for the serial concatenated coding scheme is shown in Fig. \ref{fig:systemmodel}. It consists of an outer FEC code (rate = $R_{c}^o$), a random interleaver, an inner RLL code (rate = $R_{c}^i$) and a dimming encoder (puncturing + compensation symbol insertion, rate = $R_d$). The overall rate of the concatenated scheme, $R_{c}$ = $R_{c}^o \times R_{c}^i \times R_d$.  We are using  memory two convolutional code as the FEC code because of its low complexity, flexible code rate, and soft decodability. We have assumed on-off keying (OOK) modulation \cite{b4} for LEDs where the high-level bit `1' is mapped as the LED is ON and the low-level bit `0' is mapped as the LED is OFF. 

\begin{table}[t]
\caption{Comparison of various RLL codes}
\begin{center}
\begin{tabular}{|c|c|c|c|c|c|}
\hline
\textbf{Parameter} & \textbf{Manchester} & \textbf{BMC}& \textbf{4B6B} & \textbf{Split Phase}\\
\hline
Code rate & 1/2 & 1/2 & 2/3 & 1/2\\
\hline
Iterative decoding & No & Yes & No & Yes \\
\hline
50$\%$ dimming & Yes & Yes & Yes & Yes \\
\hline
Super symbol & No & Yes & No & No \\
\hline
Run length & 2 & 2 & 4 & 2\\
\hline
\end{tabular}
\label{table:tab2}
\end{center}
\end{table}

At the transmitter, the information bit sequence, $\textbf{u}=(u_1,u_2,\dots,u_k)$ is first encoded by the outer convolutional encoder of rate $R_{c}^o=\frac{k}{n}$. Now the encoded bit sequence $\bar{\textbf{u}}=(\bar{u}_1,\bar{u}_2,\dots,\bar{u}_n)$ is passed to the interleaver ($\pi$) and become $\textbf{v}=(v_1,v_2,\dots,v_n)$. This interleaved bit sequence $\textbf{v}$ is now encoded by an inner RLL encoder of rate $R_{c}^i$ and then by dimming encoder ($d$) to become $\bar{\textbf{v}}=(\bar{v}_1,\bar{v}_2,\dots,\bar{v}_p)$. Here $\bar{\textbf{v}}$ is OOK modulated and transmitted as a VLC signal. Signal detected by a photo detector at the receiver is given by $\textbf{y}=\bar{\textbf{v}}+\textbf{n}$, where $\textbf{n}$ is the additive white Gaussian noise (AWGN) with mean zero and variance $\sigma^2$ \cite{b4}. The corresponding logarithmic likelihood ratios (LLRs) are passed to the dimming decoder followed by inner SISO RLL decoder. Finally, iterative decoding between the inner RLL decoder and the outer FEC decoder runs multiple times for exchanging extrinsic information and to refine the decoding result as shown in Fig. \ref{fig:systemmodel}.

\section{Iterative decoding $\&$ EXIT chart}

\begin{figure}[t]
\centerline{\includegraphics[scale=0.14]{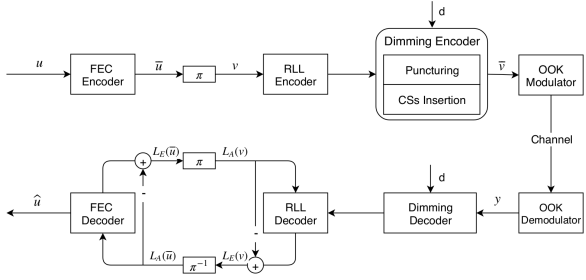}}
\caption{VLC system model}
\label{fig:systemmodel}
\end{figure}

The inner RLL decoder and the outer FEC decoder in Fig. \ref{fig:systemmodel} are both SISO decoders.
We have used the log-MAP version of BCJR decoding  algorithm \cite{b9} for both outer and inner code decoding. The iterative decoding (iteration = $1,2,\dots,L$) works in the manner that the RLL decoder takes incoming LLR $\textbf{y}=(\bar{y}_1,\bar{y}_2,\dots,\bar{y}_p)$ received from the channel and a priori L-values $L_A(\textbf{v})=(L_A(v_1),L_A(v_2),\dots,L_A(v_n))$ received from the previous iteration of FEC decoder, as input and computes the extrinsic a posteriori L-values $L_E(\textbf{v})=(L_E(v_1),L_E(v_2),\dots,L_E(v_n))$ by applying the log-MAP BCJR algorithm for all iteration = $1,2,\dots,L$ as:
{\allowdisplaybreaks
\begin{IEEEeqnarray}{lll}
L_E(v_l)&=&\ln\frac{\sum\limits_{(s',s)\in \sum_{v_l=1}}\exp(\alpha_l^*(s')+\gamma_l^*(s',s)+\beta_{l+1}^*(s))}{\sum\limits_{(s',s)\in \sum_{v_l=0}}\exp(\alpha_l^*(s')+\gamma_l^*(s',s)+\beta_{l+1}^*(s))}\notag\\ 
&& -L_A(v_l)\label{eq}
\end{IEEEeqnarray}}
where $\sum_{v_l=1}$ is set of all state pairs $s_l=s'$ and $s_{l+1}=s$ that correspond to the bit $v_l=1$ at time $l$, $\sum_{v_l=0}$ is set of all state pairs $s_l=s'$ and $s_{l+1}=s$ that correspond to the bit $v_l=0$ at time $l$, $\alpha_l^*$ is the forward metric in $\log$ domain, $\beta_l^*$ is the backward metric in $\log$ domain and $\gamma_l^*$ is the transition metric in $\log$ domain for all $l=1,2,\dots,n$. 
The calculation of $\alpha_l^*$ and $\beta_l^*$ is same as in conventional log-MAP BCJR algorithm \cite{b9} but due to OOK modulation the $\log$ domain transition metric for inner RLL decoder changes \cite{b4} and can be calculated as:
\begin{IEEEeqnarray}{lll}
\gamma_l^*(s',s)=v_lL_A(v_l)+\frac{1}{2\sigma^2}[2\textbf{y}_l\bar{\textbf{v}}_l-\textbf{y}_l^2]
\end{IEEEeqnarray}

The a posteriori L-values $L_E(\textbf{v})$ computed using (\ref{eq}) is then de-interleaved ($\pi^{-1}$) and becomes a priori L-values $L_A(\bar{\textbf{u}})=(L_A(\bar{u}_1),L_A(\bar{u}_2),\dots,L_A(\bar{u}_n))$. Now the FEC decoder takes $L_A(\bar{\textbf{u}})$ as input and computes extrinsic a posteriori L-values $L_E(\bar{\textbf{u}})=(L_E(\bar{u}_1),L_E(\bar{u}_2),\dots,L_E(\bar{u}_n))$ for all the convolutional codewords by applying the conventional BCJR algorithm for iteration = $1,2,\dots,L-1$. The $L_E(\bar{\textbf{u}})$ is then interleaved ($\pi$) and becomes a priori L-values $L_A(\textbf{v})=(L_A(v_1),L_A(v_2),\dots,L_A(v_n))$ for the next iteration of RLL decoder. The process continues till the last iteration $(=L)$ and FEC decoder computes LLR of information bits, $L(\textbf{u})=(L(u_1),L(u_2),\dots,L(u_k))$ in the last $L^{th}$ iteration and takes hard decision $\hat{\textbf{u}}$ on information bits. Some key points to note about the iterative decoding procedure are:
\begin{itemize}
\item For the first iteration of RLL decoder, the a priori L-values $L_A(\textbf{v})$ are set to all zero sequence.
\item There are two inputs to RLL decoder i.e., $\bar{\textbf{y}}$ and $L_A(\textbf{v})$ but only one input for FEC decoder i.e., $L_A(\bar{\textbf{u}})$ because there doesn't exist any value directly received from the channel for FEC decoder.
\item The FEC decoder computes extrinsic L-values for all the code bits not just the information bits.
\end{itemize}

We have also employed the EXIT chart tool \cite{ten2000design,banerjee2001asymmetric} for analyzing the exchange of information between the inner RLL decoder and the outer FEC decoder during iterative decoding. For this, we have used $R_c=\frac{1}{3}$ code with four different CC-RLL serial concatenation as:
\begin{enumerate}
\item CC-4B6B, $R_{c}^o=\frac{1}{2}$ and $R_{c}^i=\frac{2}{3}$
\item CC-Manchester, $R_{c}^o=\frac{2}{3}$ and $R_{c}^i=\frac{1}{2}$
\item CC-BMC, $R_{c}^o=\frac{2}{3}$ and $R_{c}^i=\frac{1}{2}$
\item CC-split phase, $R_{c}^o=\frac{2}{3}$ and $R_{c}^i=\frac{1}{2}$
\end{enumerate}
We used a memory two systematic convolutional code with generator matrix, $G=[1$ $\frac{5}{7}]$. As this is a rate $\frac{1}{2}$ code, we have used puncturing pattern, $P =\begin{bmatrix} 1 & 0 \\ 1 & 1 \end{bmatrix}$ to convert it  to rate $\frac{2}{3}$ code.

\begin{figure}[t]
\centerline{\includegraphics[scale=0.55]{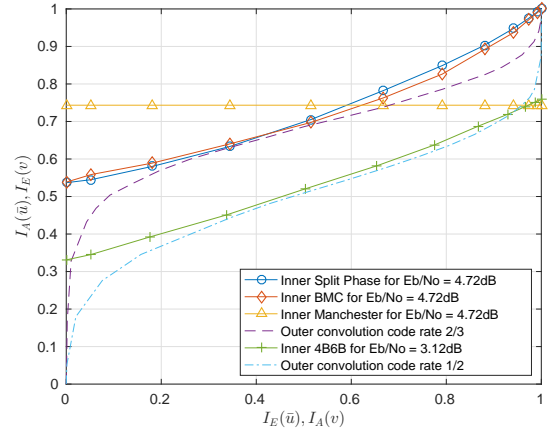}}
\caption{EXIT curve of the serial concatenated CC-RLL scheme with overall rate $\frac{1}{3}$.}
\label{fig:exit}
\end{figure}

Fig. \ref{fig:exit} shows the EXIT chart of four different CC-RLL serial concatenation explained above. On the ordinate, the inner RLL extrinsic output $I_E(v)$  becomes outer convolutional a priori input $I_A(\bar{u})$ (no changes in mutual information after interleaving). On the abscissa, outer convolutional extrinsic output $I_E(\bar{u})$ becomes inner RLL a priori input $I_A(v)$. We can make the following conclusion from the EXIT chart of Fig. \ref{fig:exit} :
\begin{itemize}
\item The convergence threshold of CC-split phase is same as that of CC-BMC concatenation ($4.72$ dB). However, CC-split phase has less decoding complexity due to its two-state structure of RLL code.
\item The concatenation of CC-Manchester code doesn't provide any iterative decoding gain.
\item The concatenation of CC-4B6B provides iterative gain initially and has a low convergence threshold value ($3.12$ dB). However, the outer EXIT curve of 4B6B is not touching the right upper corner point of (1,1) due to which this code will have higher error floor \cite{ashikhmin2004extrinsic}.
\end{itemize}

\section{Simulation Results}
\begin{figure}[t]
\centerline{\includegraphics[scale=0.61]{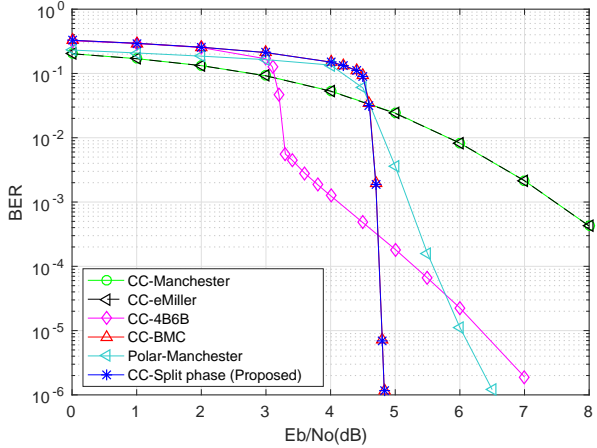}}
\caption{BER performance of proposed scheme for 50$\%$ dimming.}
\label{fig:ber}
\end{figure}

Fig. \ref{fig:ber} compares the BER performance of different RLL codes in serial concatenation with FEC codes for 50$\%$ dimming ($d=0.5$). We considered an overall rate, $R_c=\frac{1}{3}$ code with interleaver length = 32768 for all the simulations. Convolutional code parameters are same as used for EXIT chart in the previous section. There is no role of dimming encoder for 50$\%$ dimming because RLL code provides 50$\%$ dimming. The maximum number of iteration ($L$) performed is 100, with a genie stopping rule \cite{b11}. The result of Fig. \ref{fig:ber} confirms the two predictions of the EXIT chart. First, CC-4B6B concatenation have low BER in waterfall region but high BER in error floor region and second, CC-BMC and CC-split phase concatenation have same BER performance as predicted by there convergence threshold.

\begin{figure}[t]
\centerline{\includegraphics[scale=0.6]{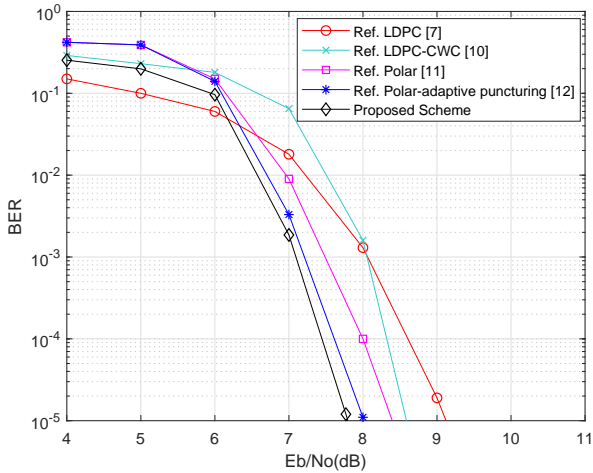}}
\caption{BER performance of proposed scheme for 60$\%$ dimming.}
\label{fig:ber1}
\end{figure}

Fig. \ref{fig:ber1} compares the BER performance of different VLC coding technique for 60$\%$ dimming ($d=0.6$). We considered an input block of length 512 bits and overall code rate of $\frac{1}{4}$ for comparison with other published results of \cite{b15},\cite{b22}-\cite{b23}. Convolutional code parameters are same as used for EXIT chart and dimming scheme used is same as in \cite{b21}. We have considered $R_c^o=1/2$, $R_c^i=1/2$ and $R_d=1$. We are achieving dimming by puncturing $p$ bits from encoded RLL code sequence and inserting $p$ bit compensation symbol sequence of all 1's.

\section{Conclusion}
In this paper, we have proposed a low complexity split phase RLL code which can be used in serial concatenation with a convolutional code in a VLC system. Split phase RLL code also provides perfect dimming ratio of $50\%$ at the output for each input bit, independent of its length. We used the EXIT chart to show that the CC-split phase concatenation provides convergence threshold, $\frac{E_b}{N_o}=4.72$ dB. The proposed scheme can also be used for dimming values other than $50\%$ by using puncturing and compensation symbols. Simulation results show that serial concatenation of split phase RLL code with convolutional code provides better performance than other proposed schemes in the literature.


\bibliographystyle{ieeetr} 
\bibliography{biblography1} 

\end{document}